\documentclass[aps,prl,reprint,superscriptaddress]{revtex4-2}
\usepackage{amsmath,amssymb}
\usepackage{physics}
\usepackage{graphicx}
\usepackage{hyperref}

\newcommand{\ack}[1]{\section*{Acknowledgements}#1}

\begin{document}
\title{When blinking helps: enhanced single-photon purity in the off states of perovskite quantum dots}

\author{Adam Olejniczak}
\affiliation{Centro de Fisica de Materiales (CFM-MPC), Donostia-San Sebastian 20018, Spain}
\affiliation{Photonic Nanomaterials, Istituto Italiano di Tecnologia, Via Morego 30, Genova 16163, Italy}

\author{Jehyeok Ryu}
\affiliation{Donostia International Physics Center (DIPC), Donostia-San Sebastian 20018, Spain}
\affiliation{Polymers and Materials: Physics, Chemistry and Technology, Chemistry Faculty, University of the Basque Country (UPV/EHU), Donostia-San Sebastian 20018, Spain}

\author{Francesco Di Stasio}
\affiliation{Photonic Nanomaterials, Istituto Italiano di Tecnologia, Via Morego 30, Genova 16163, Italy}

\author{Yuriy Rakovich}
\email{yury.rakovich@ehu.eus}
\affiliation{Donostia International Physics Center (DIPC), Donostia-San Sebastian 20018, Spain}
\affiliation{IKERBASQUE, Basque Foundation for Science, Bilbao 48013, Spain}
\affiliation{Centro de Fisica de Materiales (CFM-MPC), Donostia-San Sebastian 20018, Spain}

\author{Victor Krivenkov}
\email{victor.krivenkov@ehu.eus}
\affiliation{Polymers and Materials: Physics, Chemistry and Technology, Chemistry Faculty, University of the Basque Country (UPV/EHU), Donostia-San Sebastian 20018, Spain}
\affiliation{Centro de Fisica de Materiales (CFM-MPC), Donostia-San Sebastian 20018, Spain}

\begin{abstract}
Photoluminescence blinking typically degrades the single-photon purity of quantum dot emission. Here, we show that individual CsPbBr$_3$ perovskite quantum dots (PQDs) can exhibit the opposite behavior: low-emitting off states are dimmer and shorter-lived, yet more strongly antibunched than high-intensity on states. Using time-correlated single-photon counting combined with state-resolved second-order correlation analysis, we indentify a subset of PQDs in which the zero-delay second-order correlation, $g^{(2)}_0$, decreases upon switching from the on state to the off state, indicating imporved single-photon purity. In the most pronounced case, $g^{(2)}_0$ decreases from 0.23 in the on state to 0.08 in the off state. We attribute this conterintuitive behavior to self-trapped-exciton-mediated suppression of the biexciton--exciton cascade, which suppresses multiphoton emission more effectively than single-exciton emission. These findings reveal an unconventional blinking regime in PQDs and show that blinking does not necessarily degrade single-photon purity.
\end{abstract}

\maketitle

\section{Introduction}

Single-photon emitters are a key resource for quantum communication, computation, and metrology. \cite{Vajner2022}  A central figure of merit is the zero-delay value of the second-order correlation function, $g^{(2)}_0$, which quantifies the probability of emitting more than one photon per excitation event. Values $g^{(2)}_0 \ll 0.5$ are required for high-purity single-photon operation.\cite{Michler2000}

Colloidal semiconductor quantum dots (QDs) are attractive single-photon sources because they can be processed from solution and integrated into a wide range of photonic architectures.\cite{Olejniczak2024} Lead halide perovskite QDs (PQDs) are particularly promising: they exhibit high photoluminescence (PL) quantum yields (QYs), fast radiative rates, broad spectral tunability, and bright single-photon emission at room temperature.\cite{Park2015,Ryu2025-2} At the same time, PQDs suffer from pronounced PL intermittency (blinking) and biexciton emission. Both effects can degrade single-photon performance: blinking reduces the fraction of time the emitter is bright, whereas biexciton emission leads to multi-photon events that increase $g^{(2)}_0$.\cite{Nair2011,Li2018}

The single-exciton PL QY ($\eta_X$) is strongly influenced by nonradiative processes due to trap-assisted recombination and trion formation, while the biexciton PL QY ($\eta_{BX}$) is predominantly limited by Auger-like processes, in which the recombination energy of one electron--hole pair is transferred to another hot carrier.\cite{Nair2011,Makarov2016} Under weak pulsed excitation, when the average number of excitons per pulse $\langle N \rangle \ll 1$, the intrinsic zero-delay correlation of a single PQD can be expressed as $g^{(2)}_0 \approx \eta_{BX}/\eta_X$.\cite{Nair2011} This relation implies that any blinking mechanism that reduces $\eta_X$ without a comparable change in $\eta_{BX}$ will increase $g^{(2)}_0$ in low-emitting off states.

Several distinct blinking mechanisms have been identified in QDs.\cite{Nirmal1996,EfrosRosen1997,Frantsuzov2009,Galland2011,Yuan2018,Yang2024,Yang2025,Trinh2020,Biju2021,Gee2025,Panda2025,Ahmed2019,Olejniczak2022} In Auger (A-type) blinking, a charge carrier is trapped, and subsequent excitation produces a trion instead of a neutral exciton.\cite{EfrosRosen1997} The trion PL QY is strongly reduced by fast Auger recombination, while the charged biexciton QY is often suppressed less dramatically.\cite{Makarov2016,Kanemitsu2019,Yarita2017} In this case, the ratio $\eta_{BX}/\eta_X$ increases in off states, and $g^{(2)}_0$ is larger than in the on-charged on state, as observed in II--VI and perovskite systems.\cite{Manceau2014,Guo2025,Gee2025} In band-edge carrier (BC-type) blinking, the PL intermittency is attributed to fluctuating nonradiative rates of band-edge carriers, for example, due to activation and deactivation of shallow traps or surface states.\cite{Frantsuzov2009,Yuan2018,Seth2018,Panda2025,Ahmed2019} In the simplest BC-blinking picture, these fluctuations primarily affect $\eta_X$, whereas $\eta_{BX}$ remains controlled by intrinsic Auger processes. Often both A-type and BC-type blinking coexist within one QD.\cite{Han2020, Kim2019, Seth2018, Kline2025, Yuan2018,Guo2025,Yang2024} Both mechanisms degrade single-photon purity in the low-emitting off state \cite{Guo2025,Gee2025}. Therefore, understanding the relationship between blinking behavior and single-photon purity is critical for advancing PQDs as reliable single-photon sources.

This work presents a counterintuitive behavior in PQD blinking, where off states are not only dimmer and faster than on states, but also have better single-photon purity, with reduced $g^{(2)}_0$. This behavior is not captured by standard A-type or BC-type blinking models, in which off states are expected to increase \(g^{(2)}_0\) when \(\eta_X\) is reduced more strongly than the biexciton contribution. Instead, it suggests that, in some cases, blinking in PQDs is controlled by an additional nonradiative channel that suppresses biexciton emission more effectively than single-exciton emission.

\section{Experimental results}

CsPbBr$_3$ PQDs were prepared by ligand-assisted reprecipitation under ambient conditions, following Ref.~\cite{Ryu2025}. Post-synthesis nickel doping was applied to improve photostability, with a final Ni-to-Pb molar ratio of $\approx0.39\%$. Transmission electron microscopy reveals nearly cubic nanocrystals with an average edge length of $\sim 10$~nm (Fig.~1(a)). Ensemble PL is centered at 2.46 eV, consistent with band-edge emission (Fig.~S1(a)). The PL decay of PQD solution is well described by a biexponential function with an amplitude-weighted average lifetime of 11.5 ns (Fig.~S1(b)). Further details of the synthesis and characterization are provided in the Supplemental Material.

\begin{figure}[!ht]
  \centering
  \includegraphics[width=1.0\linewidth]{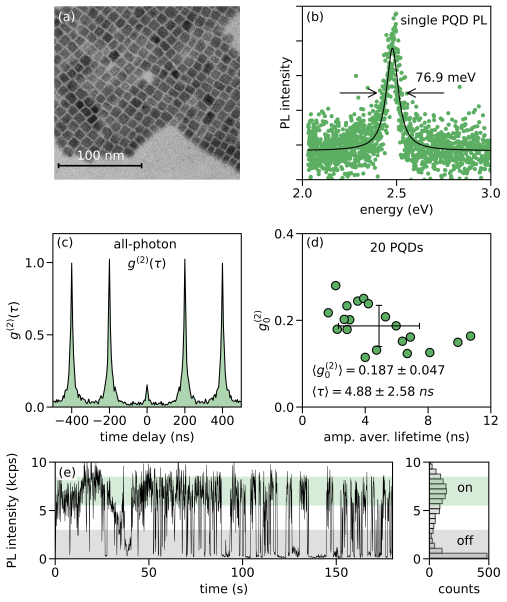}
  \caption{  
  (a) Representative TEM image of PQDs.
  (b) PL spectrum (green points) of a representative single PQD at room temperature with Lorentzian fit (black line) with 76.9 meV FWHM.
  (c) Representative all-photon second-order correlation function $g^{(2)}(\tau)$ of a single PQD. 
  (d) All-photon $g^{(2)}_0$ as a function of amplitude average PL lifetime for 20 PQDs at room temperature. Inset text and black cross are mean values with standard deviation.
  (e) Representative PL intensity trace and intensity histogram of a single PQD. Shaded regions represent high intensity (on) and low intensity (off) states.}
  \label{fig:ensemble}
\end{figure}

The optical properties were next investigated at the single-particle level. Individual PQDs exhibit Lorentzian emission spectra with a full width at half maximum (FWHM) of 76.9 meV (Fig.~1(b)). A set of individual PQDs was monitored spectrally for 300 s, and all of them showed stable emission over this interval (Fig. S2). Since the emission wavelength is closely linked to PQD size and electronic structure, the absence of spectral diffusion indicates that no significant structural changes occur on this timescale.\cite{Igarashi2023} To evaluate their single-photon emission properties, we performed time-correlated single-photon counting (TCSPC) measurements in a Hanbury Brown and Twiss configuration on 20 individual PQDs. For each PQD, second-order correlation function ($g^{(2)}(\tau)$) between two detectors was calculated for all emitted photons. 
All PQDs exhibited clear antibunching (Fig.~1(c)) with a mean single-photon purity $g^{(2)}_0$ = 0.187 and mean PL lifetime 4.88 ns (Fig.~1(d)). Thus, under the chosen excitation conditions, the investigated PQDs operate as stable room-temperature single-photon emitters with single-particle spectral width and single-photon purity comparable with previously reported state-of-the-art PQDs.\cite{Zhu2022,Park2015}

All investigated PQDs exhibit pronounced blinking between high intensity (on) and low intensity (off) states (Fig.~1(e)). To determine how blinking influences single-photon emission, we analyzed both the PL decay kinetics and the second-order correlation function separately for on and off states. (see Figs. S3-S22 in the Supplemental Material for the full datasets for all PQDs). 

State-resolved analysis revealed two qualitatively different blinking behaviors. In the first subgroup of PQDs, the off state exhibits an improvement or no change in single-photon purity, corresponding to a reduced $g^{(2)}_0$ compared to the on state. We refer to this behavior as ``single photon blinking``. The second subset of PQDs exhibits conventional behavior, in which $g^{(2)}_0$ increases in the off state, consistent with conventional A-/BC-type blinking.

A representative PQD exhibiting single-photon blinking is presented in Fig.~2(a-c). The FLID map (Fig.~2(a)) shows a linear intensity-lifetime correlation commonly associated with BC-type blinking.\cite{Yuan2018} However, state-resolved photon-correlation analysis revealed that the off state exhibits improved single-photon purity despite its reduced intensity and shortened lifetime. The $g^{(2)}_0$ value decreases by approximately a factor of 3 from on to off state, with simultaneous fivefold shortening of average lifetime (see Table I). 

\begin{figure}[!ht]
  \centering
  \includegraphics[width=1\linewidth]{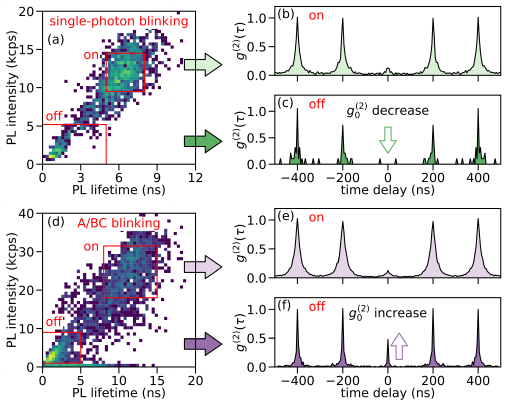}
  \caption{Distinction between single-photon blinking (a-c) and conventional A-/BC-type blinking (d-f).
  (a) FLID map of a PQD showing single-photon blinking. Red frames represent selection of bins belonging to on and off states.
  (b, c) Second-order correlation function calculated separately for on (b) and off (c) states. Green arrow in (c) points the decrease of a zero-delay peak in the off state.
  (d-f) Same data as for (a-c) but calculated for a PQD assigned to conventional A-/BC-type blinking. Violet arrow in (f) points the increase of a zero-delay peak in the off state.
  }
  \label{fig:ste_abc}
\end{figure}

A representative PQDs exhibiting a conventional blinking is shown in Fig. 2(d-f). Like the previous example, the FLID map exhibits a pronounced linear intensity-lifetime correlation. In addition, a separate region attributed to trion emission, characteristic of A-type blinking, can be distinguished.\cite{Guo2025, Yuan2018} In contrast to the previous example, single photon purity deteriorates in the off state, as expected for A-/BC-type of blinking. Here, $g^{(2)}_0$ increases nearly two times, while the average lifetime shortened fivefold (see Table I).

\begin{figure}[!ht]
  \centering
  \includegraphics[width=1.0\linewidth]{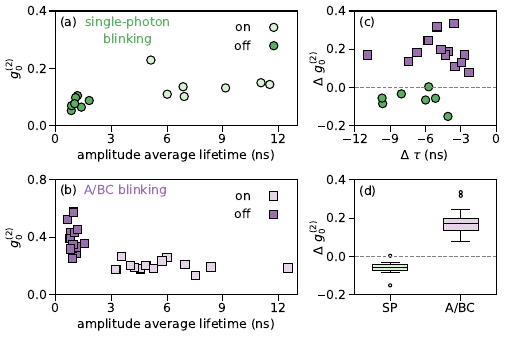}
  \caption{(a, b) State-resolved values of $g^{(2)}_{0}$ versus amplitude-weighted lifetime for (a) all PQDs assigned to single-photon blinking and (b) all PQDs assigned to conventional A-/BC-type blinking. Light- and dark-colored symbold represent values for on and off states, respectively. 
  (c) On state to off state change of $g^{(2)}_{0}$ and amplitude-weighted lifetime for PQDs showing single-photon blinking (green circles) and A-/BC-type blinking (violet squares).
  (d) Statistical distribution of the on state to off state change of $g^{(2)}_{0}$ for single-photon (SP) and A-/BC-type blinking PQDs.
  }
  \label{fig:statistics}
\end{figure}

To determine whether these two blinking behaviors are consistently observed across the full dataset, we have analyzed changes in PL lifetime ($\Delta \tau$)  and single-photon purity ($\Delta g^{(2)}_0$) upon switching from the on to the off state for all investigated PQDs. Absolute $g^{(2)}_0$ and lifetimes may exhibit significant particle-to-particle variability due to inherent heterogeneity in size, brightness or local environment. By focusing on the changes of these parameters upon switching from on to off state, each PQD effectively serves as its own reference. 
In total, we identified 7 PQDs (35\% of all analyzed in this study) exhibiting single-photon blinking (Fig. 3(a)). The remaining 13 PQDs (65\%) were classified as a conventional A-/BC-type blinking (Fig. 3(b)). While both groups show comparable reductions in PL lifetime upon switching to the off state, they differ fundamentally in the evolution of the single-photon purity (Fig. 3(c)). PQDs with single-photon blinking present consistently negative values of $\Delta g^{(2)}_0$, whereas conventional blinking PQDs exhibit positive $\Delta g^{(2)}_0$. The median changes are $\Delta g^{(2)}_0 = -0.057 $ and $+0.170$, respectively (Fig. 3(d)).
These results demonstrates that conventional blinking behaviour is consistent with established A- and BC-type blinking models, where reduced emission is accompanied by increased $g^{(2)}_0$.\cite{Manceau2014, Guo2025}  In contrast, the single-photon blinking subset shows identical intensity-lifetime correlation, but an opposite trend in single-photon purity. Therefore, state-resolved photon correlation analysis provides a method for distinguishing distinct blinking regimes in PQDs. 

\begin{table}[h]
\caption{State-resolved $g^{(2)}_0$, amplitude average lifetime and change from on to off state for two PQDs presented in Fig. 2}
\label{tab:example}
\begin{ruledtabular}
\begin{tabular}{lccc}

\multicolumn{4}{c}{\textbf{Single-photon blinking}} \\

 & on & off & $\Delta_{on \rightarrow off}$ \\
\hline
$g^{(2)}_0$ & 0.229 & 0.077 & -0.152 \\
$\tau_{avg}~(ns)$ & 5.13 & 1.04 & -4.09 \\

\hline

\multicolumn{4}{c}{\textbf{Conventional A/BC blinking}} \\

 & on & off & $\Delta_{on \rightarrow off}$ \\
\hline
$g^{(2)}_0$ & 0.180 & 0.349 & 0.169 \\
$\tau_{avg}~(ns)$ & 5.26 & 0.95 & -4.31 \\

\end{tabular}
\end{ruledtabular}
\end{table}

\section{Discussion and outlook}

The single-particle measurements on CsPbBr$_3$ PQDs reveal two qualitatively distinct regimes of PL blinking. For the A-type (trion formation) regime, PL intensity drop is accompanied by an increase of $g^{(2)}_0$, as it was previously reported for both II--VI QDs and PQDs.\cite{Manceau2014,Guo2025} In contrast some of the PQDs display a reduction of $g^{(2)}_0$ in off states, i.e., an improvement of single-photon purity. The off states are dimmer and faster than on states, yet have better single-photon purity, a behavior that cannot be explained by the conventional BC-type and A-type blinking models. In the conventional models, based on the formation of trion or short-living defect states, the reduction of $\eta_X$ should always be stronger than the reduction of $\eta_{BX}$ (see Supplemental Material for detailed explanation). Within the weak-excitation relation \(g^{(2)}_0 \approx \eta_{BX}/\eta_X\), the reduction of \(g^{(2)}_0\) in the off-state implies a lower effective \(\eta_{BX}/\eta_X\) ratio. Thus, although the off state has a reduced single-exciton emission probability, the effective biexciton contribution is suppressed even more.

We hypothesize that the nature of single-photon blinking is the formation of self-trapped excitons (STEs), which are known to arise in halide perovskites due to strong electron–phonon coupling and lattice softness, and do not require static disorder in the ground-state lattice.\cite{Fowler1973, Williams1986, Li2019, Yang2021,Ma2019,Wu2015} Electron–phonon interaction plays an important role in the optical and kinetic properties of QDs, affecting their blinking behavior \cite{Podshivaylov2023,Podshivaylov2025, Feld_NatComm_2026}. In metal–halide perovskites, STEs form via local lattice relaxation around the photoexcited state. They are highly localized and are Frenkel-like type excitons, in contrast with Wannier–Mott band-edge ones, distributed over the nanocrystal volume,\cite{Benin2018,Cortecchia2017} leading to states that are weakly emissive and long-lived at room temperature compared to the band-edge exciton.\cite{Yang2021} Because an STE is highly localized and largely decoupled from the delocalized band-edge exciton manifold, self-trapping of a hot exciton prevents the formation of a delocalized neutral biexciton during that excitation cycle.\cite{Mi2023}

It was recently proposed by Mi et al. that in some CsPbBr$_3$ PQDs the origin of blinking can also be related to STE formation.\cite{Mi2023} In our model, we assume that the intrinsic radiative rate $k_{r}^{X}$ remains essentially unchanged, whereas the total decay rate increases due to STE-related nonradiative processes. As discussed in detail in the Supplemental Material (section S5), the STE-mediated blinking mechanism provides a natural explanation for the strong suppression of biexciton emission inferred from our $g^{(2)}_0$ data.

Crucially, STE formation competes directly with biexciton ($BX$) formation at the level of hot excitons ($X'$). When two photons are absorbed, and the system is promoted from the ground state $\ket{g}$ to a state $\ket{X',X'}$ with two hot excitons, the biexciton cascade $\ket{g} \to \ket{X',X'} \to \ket{BX} \to \ket{X} \to \ket{g}$ yields two prompt band-edge photons per excitation cycle only if both hot excitons relax to the biexciton state $\ket{X',X'} \to \ket{BX}$. If instead one of the hot excitons self-traps, the system evolves into configurations such as $\ket{X,\mathrm{STE}}$ or $\ket{\mathrm{STE,STE}}$, where at most one delocalized band-edge exciton remains and the STE is long-lived, red-shifted, and largely non-emissive on the measurement timescale. Such cycles, therefore, contribute at most to a single prompt band-edge photon and effectively do not generate a biexciton–exciton two-photon cascade.

A simple model (Fig. \ref{fig:Scheme}) demonstrates how the STE-related blinking will affect single-photon purity. The model assumes that the STE is sufficiently long-lived to survive over at least one excitation period, \(T_{\mathrm{rep}}=200\) ns, so that an STE created after one pulse can interact with a band-edge exciton generated by a subsequent pulse. Each hot exciton in $\ket{X',X'}$ self-traps with probability $p_{STE}$ or relaxes to the band edge with probability $(1 - p_{STE})$. The probability to form a neutral biexciton is then $p_{\ket{X',X'} \to \ket{BX}} = (1 - p_{STE})^2$. Only this pathway can produce two prompt band-edge photons within one excitation cycle. Self-trapping therefore suppresses the biexciton--exciton cascade. At the same time, the interaction between a band-edge exciton and a long-lived STE opens an additional Auger-like nonradiative channel, reducing the emission intensity and shortening the PL lifetime.

\begin{figure}[!ht]
  \centering
  \includegraphics[width=1\linewidth]{Figure4_revision.jpg}
  \caption{Schematic illustration of the STE-based model explaining suppressed two-photon emission. Self-trapping of one hot exciton prevents neutral biexciton formation, while the remaining band-edge exciton can undergo radiative recombination or STE-assisted nonradiative loss.}
  \label{fig:Scheme}
\end{figure}

The probability that two absorbed pump photons produce two detectable band-edge photons is therefore not described by the simple Nair relation alone, but must be evaluated within the multistate kinetic scheme (see Supplemental Material for details). Kinetic Monte Carlo simulations of this multilevel scheme reproduce the experimentally observed trend: increasing the probability of self-trapping reduces the off state lifetime and intensity while suppressing the zero-delay peak. For sufficiently strong self-trapping, the calculated \(g^{(2)}_0\) of the off state becomes equal to or lower than that of the on state (Fig.~5, appendix), in agreement with the purity-enhanced blinking data. The model also suggests that cavity enhancement of the band-edge radiative rate could partially recover off state brightness without restoring the biexciton cascade and reducing single-photon purity (Fig.~6, appendix). This shows the potential application of single-photon blinking effect for cavity-coupled PQD-based quantum emitters operating at room temperature.

These results identify single-photon blinking as a distinct photophysical regime in PQDs. In this regime, lattice self-trapping reduces brightness and lifetime but suppresses the biexciton cascade even more strongly, leading to improved photon antibunching in the off-state. More broadly, the data show that blinking not always degrades quantum emission. If it is accompanied with a multiexciton suppression, it can improve single-photon purity. 

\ack{Authors acknowledge the financial support by the Department of Science, Universities and Innovation of the Basque Government (grants no. IT1526-22, PIBA\_2024\_1\_0011, and PIBA-2023-1-0007) and the IKUR Strategy;
by the Spanish Ministry of Science and Innovation (grants no. PID2022-141017OB-I00, TED2021-129457B-I00, PID2023-146442NB-I00, PID2023-147676NB-I00).}

\bibliographystyle{apsrev4-2}
\bibliography{refs}

\section{Appendix}

\subsection{Kinetic Monte Carlo simulation}

Kinetic Monte Carlo (KMC) simulations were performed to model photon emission under pulsed excitation from an emitter with competing band-edge exciton and self-trapped exciton pathways. The simulation generates a stream of photon arrival times from stochastic transitions between electronic states, from which the pulsed second-order photon correlation function, $g^{(2)}(\tau)$, is calculated directly.
The KMC model is based on the energy level scheme, where, in addition to the conventional exciton-biexciton manifold, a self-trapped exciton pathway was included (Fig.~5a). The energy-level model consists of the ground state \(\ket{g}\), the band-edge exciton \(\ket{X}\), the biexciton \(\ket{BX}\), the self-trapped exciton \(\ket{STE}\), and the exciton – self-trapped exciton configuration \(\ket{X,STE}\). Hot-exciton relaxation following optical absorption is not treated as an explicit state. This is justified because hot-exciton relaxation occurs on a much faster timescale than the nanosecond-scale photon emission dynamics relevant to the pulsed TCSPC measurement. Instead, the effect of hot-exciton relaxation is incorporated through a branching probability after absorption from the ground state.

\begin{figure}[!ht]
  \centering
  \includegraphics[width=1\linewidth]{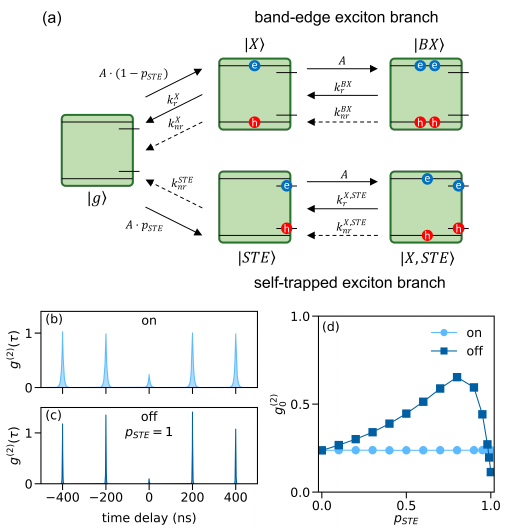}
  \caption{(a) Scheme of optical transitions included in the Kinetic Monte Carlo simulation. (b, c) Simulated second-order correlation function for the photons emitted during (b) on and (c) off state. (d) $g^{(2)}_0$ for on and off state as a function of self-trapping probability.}
  \label{fig:KMC}
\end{figure}

Upon absorption (A) from the ground state, the system populates either the self-trapped exciton pathway with probability  \(p_{\mathrm{STE}}\), or the band-edge exciton pathway with probability \(1-p_{\mathrm{STE}}\). In the on state, we set \(p_{\mathrm{STE}}=0\), whereas in the off state, \(p_{\mathrm{STE}}\) is varied to model different proabilities of self-trapped exciton formation.
Second absorption event is allowed from both band-edge and self-trapped exciton states, forming a biexciton \(\ket{X}\rightarrow \ket{BX}\), or exciton – self-trapped exciton configuration \(\ket{STE}\rightarrow \ket{X,STE}\), respectively. 
Radiative and nonradiative decay channels are included for the \(\ket{X}\), \(\ket{BX}\) and \(\ket{X,STE}\) states whereas the \(\ket{STE}\rightarrow \ket{g}\) is assumed to be purely nonradiative in the minimal model. See Supplemental Material for exact values of transition rates and more information about the KMC algorithm. 

We first investigated the evolution of the single-photon purity as a function of the self-trapped exciton branching probability, \(p_{\mathrm{STE}}\), varied between 0 and 1. For each value of \(p_{\mathrm{STE}}\), state-resolved second-order correlation function was calculated separately for photons emitted in the on and off states (Fig.~5b,c). For low and intermediate self-trapping probabilities, the off state exhibits a higher $g^{(2)}_0$ than the on state ($g^{(2)}_0 \approx 0.24$). However, when the self-trapping probability becomes sufficiently large (\(p_{\mathrm{STE}} \gtrsim 0.98\)), the trend reverses and the off state exhibits a lower $g^{(2)}_0$ than the on state (Fig.~5d), reproducing the single-photon blinking behavior observed experimentally.

\subsection{Cavity coupling}

To explore whether the reduced brightness of the off state can be compensated by enhancing its radiative emission, we next considered a PQD coupled to a cavity providing Purcell-enhanced spontaneous emission (Fig.~6a). Radiative-rate enhancement was introduced by multiplying only the radiative transition rates by the Purcell factor \(F_P\), while all nonradiative rates were kept unchanged. Simulations were performed for $F_P=10$ and $100$, which are values of Purcell factors already demonstrated experimentally for cavity coupled PQDs.\cite{Olejniczak_APLphotonics,Oddi_ACSnano, Liao_SciAdv}
The Purcell enhancement increases the QY of the off state by approximately a factor of five for $F_P=10$ and by approximately a factor of seven for $F_P=100$ (Fig.~6b). At the same time, the increased radiative rates enhance biexciton emission, leading to an increase in multiphoton events. Consequently, the on state no longer satisfies the single-photon criterion, with $g^{(2)}_0$ increasing to approximately 0.67 (Fig.~6c) and 0.85 (Fig.~6d) for $F_P=10$ and $100$, respectively. In contrast, single-photon emission can be achieved by off state for sufficiently high exciton self-trapping probabilities (\(p_{\mathrm{STE}}>0.8\) and $0.7$ for Purcell factor $10$ and $100$, respectively). Hence, the cavity-enhanced emission can substantially increase the brightness of the single-photon blinking state while preserving its antibunching characteristics (Fig.~6c,d).

\begin{figure}
  \centering
  \includegraphics[width=1\linewidth]{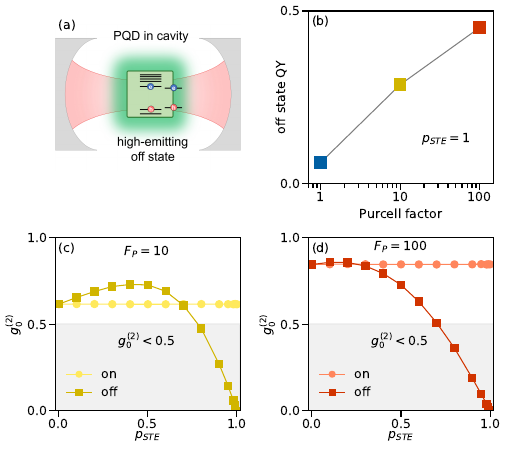}
  \caption{Effect of cavity coupling on single-photon blinking. (a) Schematic illustration of a PQD coupled to a cavity mode resulting in the increased brightess in the off state. (b) Simulated quantum yield of the off state for different values of the Purcell factor. (c,d) $g^{(2)}_0$ for on and off state as a function of self-trapping probability simulated for a PQD coupled to a cavity with Purcell factor (c) 10 and (d) 100. Shaded region in (c,d) represents a criterion of single photon emission with $g^{(2)}_0 < 0.5$}
  \label{fig:KMC_cavity}
\end{figure}

\end{document}